\pdfoutput=1
\documentclass[%
reprint,
superscriptaddress,
 amsmath,amssymb, 
 aps,
 prl,
floatfix,
]{revtex4-1}

\usepackage{graphicx}% Include figure files
\usepackage{epsfig} %Works faster than the graphicx package}
\usepackage{dcolumn}% Align table columns on decimal point
\usepackage{bm}% bold math
\usepackage{epstopdf}
\usepackage{multirow}
\usepackage{amsmath}
\usepackage{booktabs}
\usepackage[dvipsnames]{xcolor}
\usepackage{gensymb}
\usepackage{kantlipsum}  
\usepackage{hyperref}
\usepackage{ulem}
\usepackage{braket}
\usepackage{physics}
\usepackage[utf8]{inputenc}
\usepackage[T1]{fontenc}
\usepackage{mathptmx}
\usepackage{xcolor}

\begin{document}

\title{Extreme Loss Suppression and Wide Tunability of Dipolar Interactions \\ in an Ultracold Molecular Gas}

\preprint{APS/123-QED}

\author{Weijun Yuan}
\thanks{These authors contributed equally.}
\affiliation{Department of Physics, Columbia University, New York, New York 10027, USA}
\author{Siwei Zhang}
\thanks{These authors contributed equally.}
\affiliation{Department of Physics, Columbia University, New York, New York 10027, USA}
\author{Niccol\`{o} Bigagli}
\affiliation{Department of Physics, Columbia University, New York, New York 10027, USA}
\author{Haneul Kwak}
\affiliation{Department of Physics, Columbia University, New York, New York 10027, USA}
\author{Claire Warner}
\affiliation{Department of Physics, Columbia University, New York, New York 10027, USA}
\author{Tijs Karman}
\affiliation{Institute for Molecules and Materials, Radboud University, 6525 AJ Nijmegen, Netherlands}
\author{Ian Stevenson}
\affiliation{Department of Physics, Columbia University, New York, New York 10027, USA}
\author{Sebastian Will}\email{Corresponding author. Email: sebastian.will@columbia.edu}
\affiliation{Department of Physics, Columbia University, New York, New York 10027, USA}

\date{\today}

\begin{abstract}
Ultracold dipolar molecules hold great promise for the creation of novel quantum states of matter, but the realization of long-lived molecular bulk samples with strong dipole-dipole interactions has remained elusive. Here, we realize a collisionally stable gas of ultracold ground state molecules with a lifetime of several seconds. Utilizing double microwave dressing, we achieve an extreme suppression of inelastic two- and three-body losses by factors of more than 10,000 and 1,000, respectively. We find that losses remain suppressed across a wide range of dipole-dipole interactions, allowing the continuous tuning of the dipolar length from  0 to 1 \textmu m $\sim 20,000$ $a_0$. Combined with the recent realization of Bose-Einstein condensation of dipolar molecules, our findings open the door to the exploration of strongly dipolar quantum liquids.
\end{abstract}

\maketitle

\section{Introduction}

%%%%%%%%%%%%%%%%%%%%%%%%%%%%%%%%%%%%%%%%%%%%%%%%%%%%%%%%%%%%%%%%%%%%%%%%%%%%%%%%%%%%%%%%%%%%

\begin{figure*} [t]
   \centering
   \includegraphics[width=0.66\textwidth]{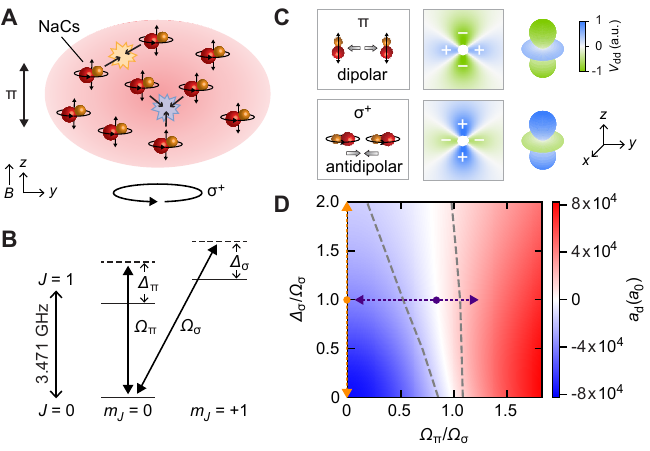}\\
   \caption{\textbf{Interaction tuning between dipolar molecules via microwave dressing.} (\textbf{A}) Illustration of microwave-dressed NaCs ground state molecules in an optical dipole trap. The simultaneous exposure to circularly and linearly polarized dressing fields can lead to a strong suppression of two-body and three-body losses. (\textbf{B}) The linearly polarized $\pi$ (circularly polarized $\sigma^{+}$) field couples $\ket{J, m_J} = \ket{0, 0}$ to $\ket{1, 0}$ ($\ket{1, 1}$) with Rabi frequency $\Omega_\pi$ ($\Omega_\sigma$) and detuning $\Delta_\pi$ ($\Delta_\sigma$). To achieve collisional shielding both fields are blue-detuned from resonance. The resonance frequency is $3.471$ GHz. (\textbf{C}) Long-range dipole-dipole interactions among molecules dressed on the $\pi$ and $\sigma^+$ transition have opposing signs, labeled as dipolar ($d_\mathrm{ind}^2 > 0$) and antidipolar ($d_\mathrm{ind}^2 < 0$), respectively. The spatial anisotropy of dipolar and antidipolar interactions corresponds to spherical harmonics $-Y_{20} \propto 1 - 3 \cos^2\theta$ and $Y_{20} \propto -(1 - 3 \cos^2\theta)$, respectively. The anisotropy is shown as a 2D cut in the $yz$ plane (middle) and in all three dimensions (right). (\textbf{D}) Dipolar length $a_{\rm d}$ for microwave dressing with $\pi$ and $\sigma^{+}$ fields as a function of  $\Omega_\pi / \Omega_\sigma$ and  $\Delta_\sigma / \Omega_\sigma$ at fixed $\Omega_\sigma / (2 \pi) = 8.1$~MHz and $(\Delta_\pi - \Delta_\sigma) / (2 \pi) = 2$~MHz. Orange and blue dots indicate the parameters for the data in Fig.~\ref{fig:2}. The dashed orange and blue arrows indicate the investigated parameter range in Fig.~\ref{fig:3} and Fig.~\ref{fig:4}, respectively.  For parameters in the area between the gray dashed lines, calculations show that the intermolecular interaction potential does not support bound states.}
   \label{fig:1}     
\end{figure*}

%%%%%%%%%%%%%%%%%%%%%%%%%%%%%%%%%%%%%%%%%%%%%%%%%%%%%%%%%%%%%%%%%%%%%%%%%%%%%%%%%%%%%%%%%%%%

Collisional stability in many-body quantum systems is an important prerequisite for the emergence of equilibrium quantum phases. Naturally occurring many-body systems such as electron gases in materials~\cite{tsui1982two, cao2018unconventional} and liquid helium~\cite{kapitza1938viscosity} are stable against inelastic loss and exhibit superconducting~\cite{bardeen1957theory} or superfluid phases~\cite{nozieres2018theory}. For quantum systems of ultracold atoms~\cite{bloch2008many}, two-body losses are intrinsically suppressed while three-body losses can be mitigated by reducing the particle density. In dilute atomic gases with interparticle spacings of 100 nm to {1~\textmu m}, collisional lifetimes reach tens of seconds, which has enabled efficient evaporative cooling and Bose-Einstein condensation~\cite{anderson1995observation, davis1995bose}. Combined with Feshbach resonances~\cite{inouye1998observation, chin2010feshbach}, which provide a high tunability of s-wave contact interactions, and optical lattices~\cite{gross2017quantum}, stable atomic quantum gases opened the field of quantum simulation, enabling the observation of BEC-BCS superfluidity~\cite{zwerger2011bcs} and the realization of strongly correlated lattice models~\cite{greiner2002quantum}. Magnetic atoms expanded ultracold quantum science towards systems with long-range dipole-dipole interactions, enabling the observation of novel superfluids~\cite{lahaye2007strong}, droplets~\cite{kadau2016observing, chomaz2016quantum}, and supersolid quantum phases~\cite{chomaz2019long, guo2019low, tanzi2019supersolid}.

Ultracold gases of dipolar molecules are emerging as an exciting new platform for many-body quantum physics~\cite{baranov2012condensed, bohn2017cold}. With a permanent electric dipole moment $d_0$ on the order of 1 Debye~(D), dipole-dipole interactions can reach a characteristic range on the micrometer scale. By controlling the alignment of the molecular dipole in the lab frame, the effective strength of dipole-dipole interactions can be tuned, promising access to new regimes of dipolar quantum matter, such as droplet formation beyond the mean field regime~\cite{ciardi2025self}, self-organized crystals~\cite{buchler2007strongly}, and lattice models with strong long-range interactions~\cite{goral2002quantum}. However, universal two-body loss~\cite{ospelkaus2010quantum, julienne2011universal}, observed in all ultracold molecular gases~\cite{bause2023ultracold}, has severely restricted sample lifetimes, posing a major impediment.

Recently, collisional shielding of molecules using static electric fields~\cite{valtolina2020dipolar, matsuda2020resonant} and microwave dressing with a single circularly polarized field~\cite{anderegg2021observation, schindewolf2022evaporation, bigagli2023collisionally, lin2023microwave} have been demonstrated. These methods have been found to suppress two-body losses by factors of 10 to 100 and enabled evaporative cooling of fermionic molecules to quantum degeneracy~\cite{valtolina2020dipolar, schindewolf2022evaporation}. In addition, double microwave dressing utilizing a combination of linearly and circularly polarized fields has been observed to strongly suppress losses in a regime of weak dipole-dipole interactions, enabling the observation of the first Bose-Einstein condensate of dipolar molecules~\cite{bigagli2023observation}. However, whether inelastic losses in molecular samples can be sufficiently suppressed while simultaneously accessing sizable dipole-dipole interactions -- essential for the realization of strongly interacting many-body systems in bulk samples of dipolar molecules -- has remained an open question.

In this work, we demonstrate simultaneous loss suppression and wide tunability of dipole-dipole interactions in a gas of ultracold NaCs ground state molecules. We show that two- and three-body losses can be reduced by factors of at least 10,000 and 1,000, respectively. For a particle density of $10^{12}$ molecules per cm$^{3}$ the observed lifetimes are greater than 6 seconds. Within the regime of extremely low losses, we find that the characteristic range of dipole-dipole interactions can be widely tuned from 0 to 1 \textmu m $\sim 20,000$~$a_0$ ($a_0$ denotes the Bohr radius). The realization of ultracold molecular gases with high collisional stability, strong long-range interactions, and flexible tunability of interactions is a critical advance that brings novel dipolar many-body systems within reach. 

\section{Interaction tuning}

Our approach to simultaneously suppress losses and achieve high tunability of dipole-dipole interactions in an ultracold molecular gas relies on double microwave dressing \cite{gorshkov2008suppression, karman2018microwave, yan2020resonant, karman2024upcoming, deng2025two}. For molecules in the rovibrational ground state, $|J, m_J \rangle = |0,0 \rangle$, blue-detuned dressing fields with $\pi$ and $\sigma^+$ polarization induce a coupling to the states $|1,0 \rangle$ and $|1,1 \rangle$, respectively, as illustrated in Figs.~\ref{fig:1}\textbf{A} and \textbf{B}. Here, $J$ is the total angular momentum of the molecule and $m_J$ its projection onto the quantization axis (in $z$ direction, set by the polarization of the dressing fields). The coupled three-level system has a collisionally-shielded eigenstate \cite{SI} $| s \rangle = \alpha \ket{0, 0} + \beta \ket{1, 0} + \gamma \ket{1, 1}$, where $\alpha$, $\beta$, and $\gamma$ denote complex probability amplitudes which are controlled by the Rabi frequencies and detunings of the dressing fields. In a classical picture, the $\pi$ and $\sigma^+$ fields induce a vertically oscillating and a horizontally rotating dipole, respectively. Fig.~\ref{fig:1}\textbf{C} illustrates the interactions for $\pi$ and $\sigma^+$ dressing at long range. For $\pi$ ($\sigma^+$) dressed molecules the interaction is repulsive (attractive) in the $xy$ plane and attractive (repulsive) along the $z$ axis. We call the interactions for $\pi$ dressing \textit{dipolar} and  \textit{antidipolar} for $\sigma^+$ dressing due to the opposite overall sign, leading to an inversion of the anisotropy of the interactions. By controlling the microwave parameters, the relative weight of dipolar and antidipolar interactions can be tuned, enabling control over the interaction strength and anisotropy.  The time-averaged dipole-dipole interaction between two dressed molecules is given by $V_{\rm dd} = d_{\rm ind}^2( 1- 3 \cos^2 \theta)/(4 \pi \epsilon_0 r^3) $. Here, $\epsilon_0$ denotes the vacuum permittivity, \textit{r} the separation between two molecules,  and $\theta$ the angle between the intermolecular and the dipolar axes (or the rotational axis of the rotating dipole). The effective dipole moment $d_{\rm ind}^2 = d_0^2 |\alpha|^2 (2|\beta|^2 - |\gamma|^2) / 3$ \footnote{The factor of 2 in front of $|\beta|^2$ arises from $\pi$ interactions being twice as strong as $\sigma^+$ interactions~\cite{karman2022resonant}. The definition of $d_{\rm ind}$ via the interaction energy formally leads to real (imaginary) values for the case of dipolar (antidipolar) interactions.} and the associated characteristic length scale, the dipolar length, $a_{\rm d} = m d_{\rm ind}^2 / (4 \pi \epsilon_0 \hbar^2 )$~\cite{buchler2007strongly,baranov2012condensed}, are determined by the composition of the shielded eigenstate. Here, $m$ denotes the molecular mass and $\hbar$ the reduced Planck constant.  

%%%%%%%%%%%%%%%%%%%%%%%%%%%%%%%%%%%%%%%%%%%%%%%%%%%%%%%%%%%%%%%%%%%%%%%%%%%%%%%%%%%%%%%%%%%%

\begin{figure*}
   \centering
   \includegraphics[width=0.66\textwidth]{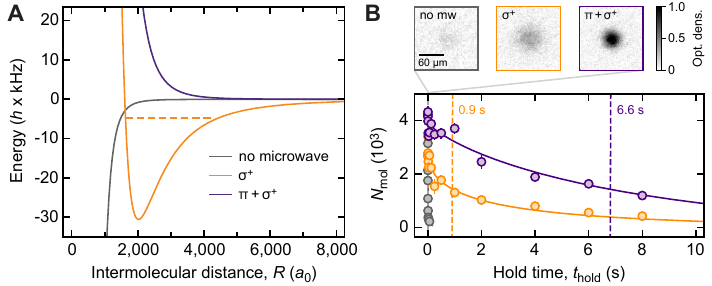}\\
   \caption{\textbf{Lifetimes of ultracold molecular samples under microwave dressing.} (\textbf{A}) Calculated intermolecular interaction potentials (s-wave channel) without microwave dressing (gray), for dressing with a $\sigma^+$ field (yellow), and for dressing with $\pi$ and $\sigma^+$ fields (purple). The parameters are $\{\Omega_\sigma, \Delta_\sigma \}/ (2 \pi) = \{ 7.9, 8\}$~MHz and $\{ \Omega_\pi, \Delta_\pi \}/ (2 \pi) = \{ 6.5, 10\}$~MHz. (\textbf{B}) (bottom) Lifetime of molecular samples without (gray), for single (yellow), and for double (purple) microwave dressing. Circles are the experimental data and solid lines are fits to the kinetic model (see Supplementary Material). Vertical dashed lines indicate the respective $1/e$ lifetimes. Error bars show the standard error of the mean from ten repetitions of the experiment. The initial sample temperature is 100(20)~nK and the initial mean density is 0.7(1)$\times10^{12}$~cm$^{-3}$. The experimental microwave parameters are $\{ \Omega_\sigma, \Delta_\sigma \}/ (2 \pi) = \{ 8.1(1), 8\}$~MHz and $\{ \Omega_\pi, \Delta_\pi \} / (2 \pi)= \{ 6.9(1), 10\}$~MHz. (top) Absorption images after $t_\mathrm{hold} = 8$ ms for the three dressing scenarios showing clouds of 630 (10), 2800(200), and 4000(200) molecules, respectively.}
   \label{fig:2}    
\end{figure*} 

%%%%%%%%%%%%%%%%%%%%%%%%%%%%%%%%%%%%%%%%%%%%%%%%%%%%%%%%%%%%%%%%%%%%%%%%%%%%%%%%%%%%%%%%%%%%

The dipolar length can be widely tuned by controlling the parameters of the dressing fields, as shown in Fig.~\ref{fig:1}\textbf{D}. The $\sigma^+$ field controls the $\ket{1, 1}$ admixture to the shielded eigenstate, inducing antidipolar interactions ($a_{\rm d} < 0$), whose strength can be tuned via $\Delta_\sigma / \Omega_\sigma$. The presence of the $\pi$ field leads to the additional admixture of $\ket{1, 0}$, inducing dipolar interactions ($a_{\rm d} > 0$). The ratio $\Omega_\pi / \Omega_\sigma$ controls the balance of dipolar and antidipolar contributions to the overall interactions. For $|\gamma|^2 = 2 |\beta|^2$ dipolar and antidipolar contributions compensate each other and long-range dipole-dipole interactions effectively vanish, $a_{\rm d} = 0$, which facilitated the Bose-Einstein condensation of NaCs~\cite{bigagli2023observation}. For $\Omega_\pi / \Omega_\sigma$ below (above) compensation the overall interactions are antidipolar (dipolar) with $a_{\rm d} < 0$ ($a_{\rm d} > 0$), with a tuning range for the dipolar length from 0 to several 10,000 $a_0$. How much of this tuning range is usable for the investigation of many-body bulk phases critically depends on the inelastic loss dynamics, which we explore in the following.

\section{Long Sample Lifetime}
Besides inducing dipole-dipole interactions, the dressing fields allow the engineering of a collisional barrier at short range, suppressing inelastic collisions between the molecules. Fig.~\ref{fig:2}\textbf{A} shows the interaction potentials of NaCs ground state molecules for three relevant dressing scenarios. Without microwave dressing, dipolar interactions are absent and short-range interactions are strongly attractive, inducing universal two-body loss ~\cite{ospelkaus2010quantum,julienne2011universal,bause2023ultracold}. For single microwave dressing with $\Delta_\sigma/\Omega_\sigma = 1$ and $\Omega_\pi/\Omega_\sigma = 0$, the dipolar length is $a_{\rm d} \sim -40,000$~$a_0$ and molecules approaching each other initially experience an attractive potential before reaching a steep repulsive barrier at an intermolecular distance $\sim 1,500$ $a_0$. For double microwave dressing with $\Delta_\sigma/\Omega_\sigma = 1$ and $\Omega_\pi/\Omega_\sigma = 0.85$, the potential has a flat tail at long range, corresponding to compensated dipolar interactions, $a_{\rm d} \sim 0$, and is purely repulsive with a steep barrier at $\sim 2,500$ $a_0$.

We experimentally probe the lifetime of the molecular gas in all three dressing scenarios. To this end, we prepare samples of $4.0(5) \times 10^3$ NaCs molecules in the rovibrational ground state at a temperature of $T = 100(20)$~nK and a mean density of $n = 0.7(1) \times 10^{12}$~cm$^{-3}$, held in a crossed optical dipole trap ($1064$ nm). By adiabatically ramping up the dressing fields, the molecules are prepared in a shielded eigenstate (for additional details see Supplementary Material and earlier work~\cite{stevenson2023ultracold, bigagli2023collisionally}). To record the decay dynamics, samples are held in trap for variable hold times before the molecule number is measured via absorption imaging after time of flight. The trap depth is chosen to be large, corresponding to a truncation parameter of  20, i.e.~the trap is 20-times deeper than the mean kinetic energy of the gas ($\sim k_\mathrm{B}T$, where $k_\mathrm{B}$ denotes Boltzmann's constant). This suppresses free evaporation as a spurious loss process which otherwise can mask more fundamental loss processes~\cite{weber2003three}. Owing to one-body loss that is caused by the scattering of trap laser light, the maximum lifetime of molecular samples in our setup is found to be $\tau_\mathrm{1B} = 8.5(1.5)$ s ~\cite{SI}. Fig.~\ref{fig:2}\textbf{B} shows the decay dynamics for the three dressing scenarios. Without microwave dressing, the decay is extremely rapid with a $1/e$ lifetime of 3.0(3) ms; with single microwave dressing, the lifetime is extended to 0.9(1) s; with double microwave dressing there is an additional significant enhancement of the lifetime to 6.6(5) s. This value is close to the maximum lifetime $\tau_\mathrm{1B}$, indicating an extreme suppression of collisional losses.

%%%%%%%%%%%%%%%%%%%%%%%%%%%%%%%%%%%%%%%%%%%%%%%%%%%%%%%%%%%%%%%%%%%%%%%%%%%%%%%%%%%%%%%%%%%%

\begin{figure*}
    \centering
    \includegraphics[width=0.67\textwidth]{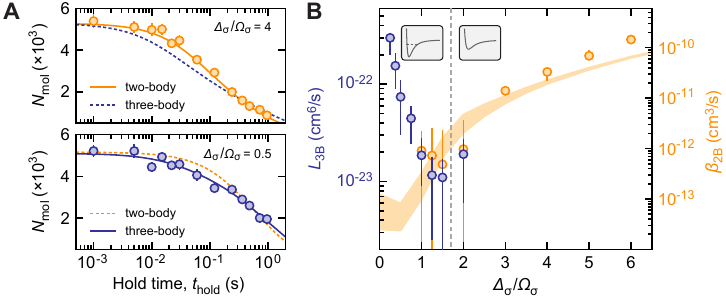}
    \caption{\textbf{Precision measurement of molecular two- and three-body loss for single microwave dressing.} (\textbf{A}) Decay dynamics at $\Delta_{\sigma} / \Omega_{\sigma} = 4$ (top) and $\Delta_{\sigma} / \Omega_{\sigma} = 0.5$ (bottom). Circles are experimental data, yellow (blue) lines show the fit of a two-body (three-body) decay model; solid (dashed) lines indicate the fit with lower (higher) $\chi^2$ value. Error bars show the standard error of the mean from ten repetitions of the experiment. (\textbf{B}) Measured two- and three-body loss rate coefficients, $\beta_\mathrm{2B}$ and $L_\mathrm{3B}$, as a function of $\Delta_{\sigma}/\Omega_{\sigma}$ obtained from fits that simultaneously take into account two- and three-body loss. Yellow (blue) circles show $\beta_\mathrm{2B}$ ($L_\mathrm{3B}$). The shaded yellow region shows the results of a two-body coupled-channels calculation for 3(2) degrees ellipticity of the $\sigma^+$ microwave field. The dashed vertical line indicates the appearance of a bound state in the collisional potential for $\Delta_\mathrm{\sigma}/ \Omega_\sigma < 1.7$ (calculated), also illustrated in the insets. Error bars show the $1\sigma$ uncertainty from fits of the loss model (see Supplemental Material). Data in this figure is recorded with molecular samples at an initial temperature of 250(50)~nK, an initial mean density of $0.4(1) \times 10^{12}$~cm$^{-3}$, $\Omega_{\sigma} / (2 \pi) = 4.0(1)$~MHz, and a measured one-body lifetime of $1/\Gamma_{\mathrm{1B}} = 3.0(5)$ s.}
    \label{fig:3}
\end{figure*}

%%%%%%%%%%%%%%%%%%%%%%%%%%%%%%%%%%%%%%%%%%%%%%%%%%%%%%%%%%%%%%%%%%%%%%%%%%%%%%%%%%%%%%%%%%%%

\section{Precision Analysis of Losses}

Analyzing the losses in greater detail, we first consider the case of single microwave dressing, tuning $\Delta_\sigma/\Omega_\sigma$, while $\Omega_\pi/\Omega_\sigma = 0$ stays constant (see Fig.~\ref{fig:3}). Given a sufficient signal-to-noise ratio, the measured decay curves allow us to extract the contributions of two- and three-body loss, due to their characteristic density dependence, proportional to $n$ and $n^2$, respectively. Fig.~\ref{fig:3}\textbf{A} shows two exemplary curves for large ($\Delta_\sigma / \Omega_\sigma = 4$) and small ($\Delta_\sigma / \Omega_\sigma = 0.5$) detunings, revealing a noticeable difference in the decay dynamics for two- and three-body loss. By fitting the decay curves with a model that simultaneously accounts for two- and three-body loss, we can precisely extract the respective loss rate coefficients, $\beta_\mathrm{2B}$ and $\Gamma_\mathrm{3B}$ (details are provided in the Supplementary Material). 

Fig.~\ref{fig:3}\textbf{B} summarizes the measured two- and three-body loss rate coefficients as a function of $\Delta_\sigma/\Omega_\sigma$. For large detunings, $\Delta_{\sigma}  / \Omega_{\sigma}  > 2$, we observe pure two-body loss, in excellent agreement with our two-body coupled-channels calculation, while three-body loss is not detectable. The range $1 < \Delta_{\sigma}  / \Omega_{\sigma}  < 2$ is a transition region with significant contributions of both two- and three-body loss. For $\Delta_{\sigma}  / \Omega_{\sigma}  < 1$, we observe dominant three-body loss that steeply increases for smaller $\Delta_\sigma$, originating from three-body recombination when a bound state is present in the microwave-dressed interaction potential~\cite{avdeenkov2003linking, chen2024ultracold} as discussed in recent theoretical work~\cite{stevenson2024three}. In the trade-off between two- and three-body loss, the optimal detuning of single microwave dressing is $\Delta_\sigma/ \Omega_\sigma \sim 1$, where losses are suppressed by about a factor of 300. 

%%%%%%%%%%%%%%%%%%%%%%%%%%%%%%%%%%%%%%%%%%%%%%%%%%%%%%%%%%%%%%%%%%%%%%%%%%%%%%%%%%%%%%%%%%%%

\begin{figure} [t]
    \centering
    \includegraphics{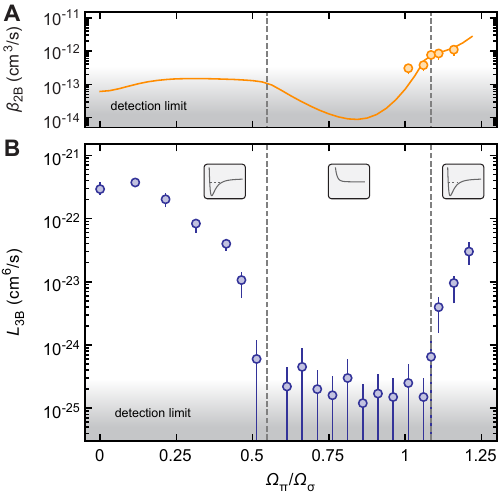}
    \caption{\textbf{Extreme suppression of collisional loss via double microwave dressing.} (\textbf{A}) Measured two-body loss rate coefficient $\beta_\mathrm{2B}$ as a function of $\Omega_\mathrm{\pi} / \Omega_\sigma$ (yellow circles) and coupled-channels calculation (yellow line)~\cite{karman2024upcoming}. (\textbf{B}) Measured three-body loss rate coefficient $L_\mathrm{3B}$ as a function of $\Omega_\mathrm{\pi} / \Omega_\sigma$. Error bars show the 1$\sigma$ error from the fit of the decay curves. The gray-shaded areas indicate the detection limit arising from the finite one-body lifetime. The dashed vertical lines indicate the appearance of bound states in the collisional potential for $\Omega_\mathrm{\pi}/ \Omega_\sigma < 0.55$ and $\Omega_\mathrm{\pi}/ \Omega_\sigma > 1.1$ (calculated). Data in this figure is recorded with molecular samples at an initial temperature of $100(20)$~nK, an initial mean density of $1.1(4) \times 10^{12}$~cm$^{-3}$, $\{ \Omega_\sigma, \Delta_\sigma \} / (2 \pi) = \{ 8.1(1), 8 \}~$MHz, and $\Delta_\pi / (2 \pi) = 10 $~MHz.}
    \label{fig:4}
\end{figure}

%%%%%%%%%%%%%%%%%%%%%%%%%%%%%%%%%%%%%%%%%%%%%%%%%%%%%%%%%%%%%%%%%%%%%%%%%%%%%%%%%%%%%%%%%%%%

\begin{figure}
   \centering
\includegraphics{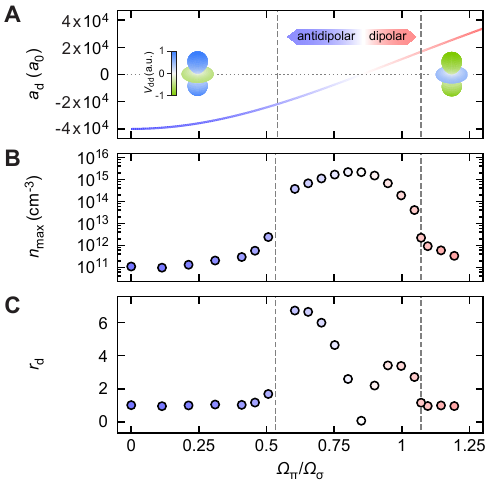}
   \caption{\textbf{Wide experimental tuning range of dipolar interactions.} (\textbf{A}) Dipolar length as a function of $\Omega_\pi / \Omega_\sigma$ at fixed $\Delta_\sigma / \Omega_\sigma = 1$. The dotted horizontal line indicates $a_{\rm d} = 0$. Insets illustrate the anisotropy of the dipolar and antidipolar interaction energy, respectively. (\textbf{B}) Maximal density $n_\mathrm{max}$ to reach a sample lifetime of 100 ms in the presence of the measured two- and three-body losses (see Supplementary Material~\cite{SI}). (\textbf{C}) Calculated ratio of dipolar and kinetic energy, $r_\mathrm{d}$, assuming a degenerate gas of NaCs molecules at density $n_\mathrm{max}$.
   }
   \label{fig:5}
\end{figure}

%%%%%%%%%%%%%%%%%%%%%%%%%%%%%%%%%%%%%%%%%%%%%%%%%%%%%%%%%%%%%%%%%%%%%%%%%%%%%%%%%%%%%%%%%%%%

\section{Extremely Low Losses}

Now, we analyze the decay dynamics for the case of double microwave dressing. While keeping the $\sigma^+$ field constant at $\Delta_\sigma/ \Omega_\sigma = 1$, we add the $\pi$ field and explore the loss processes in the experimentally accessible range $0 < \Omega_\pi / \Omega_\sigma <1.2$. Initially, the samples are prepared under compensated conditions ($\Omega_\pi / \Omega_\sigma = 0.85$), followed by a slow ramp of $\Omega_\pi / \Omega_\sigma$ to the target value, where the decay curves are recorded and analyzed following the fitting method described above. Additional details are provided in the Supplementary Material. Across a wide range of $\Omega_\pi / \Omega_\sigma$ values, we observe an extreme suppression of losses as shown in Fig.~\ref{fig:4}. Due to the maximum lifetime $\tau_\mathrm{1B}$, we cannot detect two- and three-body loss coefficients below $10^{-13}$~cm$^3$/s and $1 \times 10^{-25}$~cm$^6$/s, respectively, for a typical density of $10^{12}$ cm$^{-3}$. Indeed, we observe that both $\beta_\mathrm{2B}$ and $\Gamma_\mathrm{3B}$ can be suppressed below this limit. 

Two-body loss is only detectable for $\Omega_\pi / \Omega_\sigma > 1$ and the measured values are in  excellent agreement with our coupled-channels calculation (see Fig.~\ref{fig:4}\textbf{A}). In the range $0 < \Omega_\pi / \Omega_\sigma < 1$ two-body loss is suppressed by at least a factor 10,000 compared to the two-body loss rate coefficient of $10^{-9}$~cm$^3$/s in the absence of microwave dressing. To extract the three-body loss rate coefficients, we now fix the two-body coefficients in our fitting model to the theoretical values, due to the good agreement in the regimes where experimental and theoretical values can be compared. For $0.6 < \Omega_\pi / \Omega_\sigma < 1.1$, we measure three-body loss rate coefficients of $2(2) \times 10^{-25}$~cm$^6$/s, at or below our detection limit (see Fig.~\ref{fig:4}\textbf{B}). Outside of this stability window three-body losses rise rapidly. This rise coincides with the appearance of bound states in the microwave-dressed interaction potential, as indicated by the dashed lines in Figs.~\ref{fig:1}\textbf{D} and Figs.~\ref{fig:4}, opening a channel for three-body recombination~\cite{stevenson2024three}. Compared with $\Gamma_\mathrm{3B} = 3(1) \times 10^{-22}$~cm$^6$/s under single microwave dressing at $\Omega_\pi/\Omega_\sigma = 0$, double microwave dressing can suppress three-body losses by at least a factor of 1,000.

\section{Access to strong interactions}

The high collisional stability across a wide range of microwave parameters comes with a wide tuning range of dipolar and antidipolar interactions in our system. Within $0.6 < \Omega_\pi / \Omega_\sigma < 1.1$, the dipolar length can be continuously tuned from $-21,000\, a_0$ to $12,000\, a_0$ (see Fig.~\ref{fig:5}\textbf{A}). 

The ratio $r_\mathrm{d}$ of dipolar interaction energy, $E_\mathrm{d}$, and kinetic energy, $E_\mathrm{kin}$, is a helpful metric to assess the interaction strength in a dipolar gas~\cite{buchler2007strongly,baranov2012condensed}. For a degenerate bulk gas in three dimensions, $E_\mathrm{d} = |d_\mathrm{ind}|^2 n/ (4 \pi \epsilon_0) = ( \hbar^2/m) |a_\mathrm{d}| n$ and $E_\mathrm{kin} = (\hbar^2/m) n^{2/3}$. Noting that the mean interparticle spacing is related to the density via $a_\mathrm{sp} = n^{-1/3}$, the interaction strength is given by $r_\mathrm{d} = |a_\mathrm{d}|/a_\mathrm{sp}$ and  $r_\mathrm{d}>1$ indicates that dipolar interactions dominate over kinetic energy. To assess which densities are realistically achievable, we consider the measured two- and three-body loss rate coefficients of our system. For a target sample lifetime of 100 ms, which will be sufficient to achieve thermal equilibrium under typical conditions, we calculate the maximal allowable density $n_\mathrm{max}$ as a function of $\Omega_\pi / \Omega_\sigma$, see Fig.~\ref{fig:5}\textbf{B}. Remarkably, the highest densities are in the $\sim 10^{15}$ cm$^{-3}$ range, similar to the highest densities in ultracold atomic samples. Assuming the calculated maximal densities, $r_\mathrm{d}$ covers the range from 0 to 6.5, as shown in Fig.~\ref{fig:5}\textbf{C}, allowing the continuous tuning from weakly to strongly dipolar quantum systems.  

\section{Discussion and Outlook}

In this work, we have demonstrated the extreme suppression of two- and three-body losses in an ultracold gas of bosonic NaCs molecules, while simultaneously enabling a wide tunability of dipole-dipole interactions. Similar to the impact of Feshbach resonances for atomic quantum gases, we anticipate that  microwave dressing will play a critical role for the investigation of many-body physics with molecular quantum gases. The method should be applicable to a broad range of molecular species with sufficiently large dipole moments~\cite{karman2024upcoming, narb2025BEC}. Potentially, the stability window and the associated range of dipolar lengths can be further extended by adding additional dressing fields or a static electric field, which can admix higher rotational states to the shielded eigenstate. In addition, the versatile control over microwave dressing fields allows new types of quenching experiments and Floquet engineering of dipolar interactions.

Combined with the recently realized Bose-Einstein condensate of dipolar NaCs molecules, this work constitutes a critical advance towards the exploration of strongly dipolar quantum matter with ultracold molecules. Leveraging dipolar interactions that are about two orders of magnitude enhanced compared to magnetic atoms, new possibilities include the realization of quantum droplets beyond the mean-field regime~\cite{langen2025dipolar, deng2025two, zhang2025quantum}, exotic droplet arrays~\cite{ciardi2025self}, supersolid phases~\cite{mitra2009hexatic,cinti2014defect}, and self-organized dipolar crystals~\cite{buchler2007strongly}, as well as strongly correlated systems with unusual thermodynamic properties~\cite{sanchez2023heating, zhang2025quantum}.

\hspace{6pt}

{\bf Acknowledgments.} 
We acknowledge helpful discussions with Chris Greene,  Ahmed Elkamshishy, and Shayamal Singh. We thank Lin Su, Tarik Yefsah, and Martin Zwierlein for critical reading of the manuscript. We are grateful to Aden Lam for important contributions in the construction of the experimental apparatus. This work was supported by an NSF CAREER Award (Award No.~1848466), an NSF Single Investigator Award (Award No.~2409747), an ONR DURIP Award (Award No.~N00014-21-1-2721), an AFOSR Single Investigator Award (Award No.~FA9550-25-1-0048), and a grant from the Gordon and Betty Moore Foundation (Award No.~GBMF12340). C.W.~acknowledges support from the Natural Sciences and Engineering Research Council of Canada (NSERC). I.S.~was supported by the Ernest Kempton Adams Fund. T.K.~acknowledges NWO VIDI (Grant ID 10.61686/AKJWK33335). S.W.~acknowledges additional support from the Alfred P. Sloan Foundation.

\setcounter{figure}{0}
\makeatletter 
\renewcommand{\thefigure}{S\@arabic\c@figure}
\makeatother

{\bf Data Availability.} The experimental data that supports the findings of this study are available from the corresponding author upon reasonable request. Source data are provided with this paper.

{\bf Code Availability.} All relevant codes are available from the corresponding author upon reasonable request.

% \bibliography{Literature}
%merlin.mbs apsrev4-1.bst 2010-07-25 4.21a (PWD, AO, DPC) hacked
%Control: key (0)
%Control: author (8) initials jnrlst
%Control: editor formatted (1) identically to author
%Control: production of article title (-1) disabled
%Control: page (0) single
%Control: year (1) truncated
%Control: production of eprint (0) enabled
%
\end{document}

\title{Supplementary Material for ``Extreme Loss Suppression and Wide Tunability of Dipolar Interactions in an Ultracold Molecular Gas''}

\preprint{APS/123-QED}

\author{Weijun Yuan}
\thanks{These authors contributed equally.}
\affiliation{Department of Physics, Columbia University, New York, New York 10027, USA}
\author{Siwei Zhang}
\thanks{These authors contributed equally.}
\affiliation{Department of Physics, Columbia University, New York, New York 10027, USA}
\author{Niccol\`{o} Bigagli}
\affiliation{Department of Physics, Columbia University, New York, New York 10027, USA}
\author{Haneul Kwak}
\affiliation{Department of Physics, Columbia University, New York, New York 10027, USA}
\author{Claire Warner}
\affiliation{Department of Physics, Columbia University, New York, New York 10027, USA}
\author{Tijs Karman}
\affiliation{Institute for Molecules and Materials, Radboud University, 6525 AJ Nijmegen, Netherlands}
\author{Ian Stevenson}
\affiliation{Department of Physics, Columbia University, New York, New York 10027, USA}
\author{Sebastian Will}\email{Corresponding author. Email: sebastian.will@columbia.edu}
\affiliation{Department of Physics, Columbia University, New York, New York 10027, USA}

\date{\today}

\maketitle

\setcounter{figure}{0}
\makeatletter 
\renewcommand{\thefigure}{S\@arabic\c@figure}
\makeatother

%\blue{data for fig 2 and 4 is 100 nK, 0.85e12 cm-3 and 1.1e12 cm-3 mean density. data for fig 3 is 250 nK and 0.3e12 cm-3 mean density} 

%\red{When and how is microwave switched on in the case of single sheilding?} \blue{After preparing atoms in deep trap, we make the molecules and ramp on the microwave shielding. The microwave is first ramp on at 4MHz Rabi and 12 MHz detuning in 60 us. Then we do the frequency switch in 500 us to different frequencies. This is done by switch the source from DDS to Rohde Schwarz}

\begin{figure}
    \centering
    \includegraphics{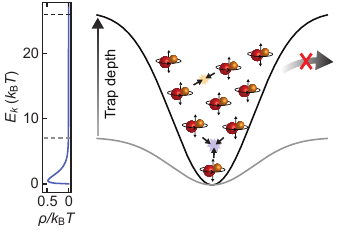}
    \caption{\textbf{Illustration of the deep trap technique.} The trap depth is ramped up after preparing a cold molecular sample to suppress evaporative loss as a spurious process, signficiantly enhancing our sensitivity to precisely detect fundamental two- and three-body loss. The left panel shows a typical Maxwell-Boltzmann distribution of molecules in a harmonic trap at $T = 150$ nK. $\rho$ is the probability distribution.}
    \label{fig:s1}
\end{figure}

\begin{figure}
    \centering
    \includegraphics{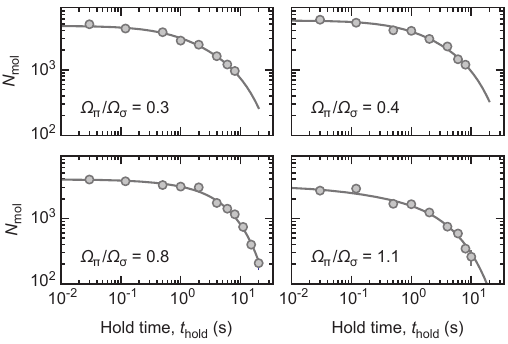}
    \caption{\textbf{Decay curves under double microwave dressing.} The respective $\Omega_\pi/\Omega_\sigma$ values are shown as an inset. $\{ \Omega_\sigma, \Delta_\sigma \} / (2 \pi) = \{ 8.1, 8\}~$MHz and $\Delta_\pi / (2 \pi) = 10$~MHz are constant for all data sets. Each data point is the average of 10 experimental runs. Solid lines show the fit from the loss model including two-body and three-body loss. Data sets with $\Omega_\mathrm{\pi}/\Omega_\mathrm{\sigma} = 0.3$ and 0.4 are taken without the vertical dipole trap (mean trap frequency $\bar{\omega}/(2 \pi)$ = 39~Hz). Data sets with $\Omega_\mathrm{\pi}/\Omega_\mathrm{\sigma} = 0.8$ and 1.1 are taken with the vertical dipole trap (mean trap frequency $\bar{\omega}/(2 \pi)$ = 70~Hz). }
    \label{fig:s2}
\end{figure}

\begin{figure}
    \centering
    \includegraphics{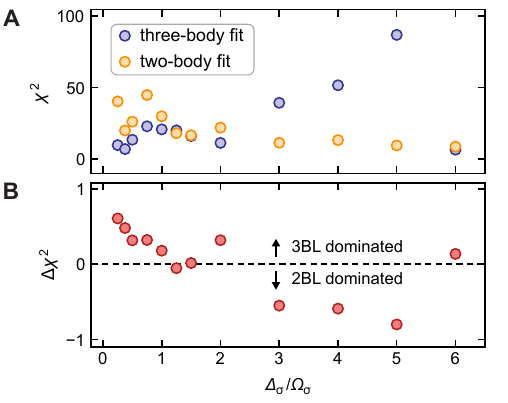}
    \caption{\textbf{Analysis of two- and three-body fit models for single microwave dressing.} (\textbf{A}) $\chi^2$ of the two-body fit  ($\chi^2_\text{2B}$) versus the three-body fit ($\chi^2_\text{3B}$) for different $\Delta_\mathrm{\sigma} \Omega_\sigma$. For this data $\Omega_\mathrm{\sigma}/(2 \pi) = 4$~MHz. (\textbf{B}) Contrast $\Delta \chi^2 = (\chi^2_\mathrm{2B} - \chi^2_\mathrm{3B})/(\chi^2_\mathrm{2B}+\chi^2_\mathrm{3B})$ between the two- and three-model fits.}
    \label{fig:s3}
\end{figure}

{\bf Preparing molecules in the deep trap.} 
For the experiments on single microwave dressing, the deep trap method (see Fig.~\ref{fig:s1}) is implemented as follows: Before associating molecules, the atomic mixture is cooled to 150~nK in a crossed optical dipole trap formed by two horizontally crossed laser beams at a wavelenght of $\lambda = 1064$ nm. Then, the trap is ramped from a depth of 1~\textmu K to 4~\textmu K within 25~ms right before molecular association. NaCs Feshbach molecules are created from Na and Cs atoms via magneto-association~\cite{lam2022high}, followed by stimulated Raman adiabatic passage (STIRAP) to the rovibrational ground state~\cite{cairncross2021assembly,stevenson2023ultracold, warner2023pathway}. The permanent dipole moment of NaCs ground state molecules is 4.6(2) D ~\cite{dagdigian1972molecular,aymar2005calculation}. The initial temperature of NaCs ground state molecules is 250(50)~nK; the corresponding trap truncation parameter is $U_{\mathrm{trap}}/k_{\mathrm{B}}T \sim$~19. The trap frequencies are $\omega/(\mathrm{2 \pi}) = (56,49,128)$~Hz and the mean density is $0.3 \times 10^{12}$~cm$^{-3}$. Then the $\sigma^{+}$-polarized microwave field is ramped to $\{\Omega_\mathrm{\sigma},\Delta_\mathrm{\sigma}\}/(2 \pi) = \{4,12\}$~MHz within 60~\textmu s. The molecules are held for 15~ms after which the microwave frequency is switched to the final value within 100~\textmu s, marking the beginning of the loss measurement. 

For the experiments on double microwave dressing, molecular evaporation enables a more direct way to realize the deep trap method: First, ground state molecules are created with a temperature of 750(50)~nK. Then double microwave shielding at the compensation point, $\{ \Omega_\sigma, \Delta_\sigma \}/ (2 \pi) = \{ 8.1(1), 8\}$~MHz and $\{ \Omega_\pi, \Delta_\pi \} / (2 \pi)= \{ 6.9(1), 10\}$~MHz, is turned on and the molecules are evaporatively cooled to 50~nK under ideal conditions. Next, a third vertically oriented dipole trap is turned on within 100~ms to increase the mean trap frequency from $\bar{\omega} / (2 \pi) = 39$~Hz to $\bar{\omega} / (2 \pi) = 60$~Hz. After holding for 100~ms to allow for rethermalization, we recompress the dipole trap to 2.5~\textmu K and tune $\Omega_{\pi}$ to the target value in 100 ms.  The vertical dipole trap increases the mean density to $>1 \times 10^{12}$~cm$^{-3}$ increasing our sensitivity to collisional loss. The temperature of the molecular cloud is 75(15)~nK and the truncation parameter is $U_{\mathrm{trap}}/k_{\mathrm{B}}T \sim 20$. The trap frequencies are $\omega/\mathrm{2 \pi} = (49,53,133)$~Hz with a mean trap frequency of $\bar{\omega} / (2 \pi) = 70$~Hz.

For the data in Fig.~2, a quench-type preparation sequence is utilized to prepare samples with similar initial densities: First, ground state molecules are prepared at a temperature of 750(50)~nK protected by double microwave dressing at the compensation point. Then they are evaporatively cooled to 50~nK. Next, the vertical dipole trap is turned on within 100~ms. After 100~ms wait time for rethermalization, the cloud is compressed to the deep trap configuration within 100~ms. To realize the three dressing scenarios mentioned in the main text, we either directly measure the lifetime or quench the shielding by switching off the $\pi$-field or completely switch off both the $\pi$- and the $\sigma$-field within 40~\textmu s. Importantly, both microwave dressing fields are turned on again during time-of-flight to prevent loss.

{\bf Microwave generation.}
The antenna system for microwave delivery has three paths: the $\sigma^+$ path, the $\pi$ path, and the $\pi$ compensation path (see Fig.~\ref{fig:s4}).

The $\sigma^+$ field is generated by feeding a microwave signal to the cloverleaf antenna array that is described in recent work~\cite{yuan2023planar}. The block diagram of the microwave system is shown in Fig.~\ref{fig:s4}\textbf{A}. The microwave signal is generated by a SG12000 RF generator from DS Instruments and its amplitude is controlled by a voltage controlled attenuator (General Microwave, D1954). The signal is split into two branches. Each branch is amplified by an RF amplifier (RF Bay, JPA-1000-8000-5). A narrow-band filter cavity (WT Microwave, WT-A10140-Q04) is used to suppress phase noise from the microwave source and amplifier. The signal is split and fed into one of the four ports of the antenna array. The phase of each input to the array is independently controlled by phase shifters.

The $\pi$ field is generated with a single-loop antenna oriented along the y-axis shown in Fig.~1A. The block diagram of the $\pi$ branch, as shown in Fig.~\ref{fig:s4}\textbf{B}. The microwave signal is generated with a radio frequency generator (Rohde $\&$ Schwarz SMA100B). It is split into two paths: the main $\pi$ path ($\pi_\mathrm{x}$) and the compensation path ($\pi_\mathrm{c}$). On the $\pi_\mathrm{x}$ path, the signal is amplified and passes through a filter cavity followed by a single-loop antenna. The compensation path is used to ensure that the $\pi$ field and the plane of the $\sigma^+$ field are normal to each other, compensating unwanted finite projections of the $\pi$ field onto the plane of the $\sigma^+$ field to a high degree. In the $\pi_\mathrm{c}$ path, the signal passes through a phase shifter which adjusts the relative phase between the two $\pi_\mathrm{c}$ fields to enable cancellation. To get complete freedom to cancel the projection of $\pi_x$, we feed $\pi_{\rm c}$ into the cloverleaf array through directional radio frequency couplers.

All three paths also have radio frequency switches for fast switching (Mini-Circuits, ZFSWA2R-63DR+) and isolators (Raditek, RADI-3.4-3.6-S3-10WR-10WFWD-H21) on either side of the filter cavities to avoid damage to other equipment. They are omitted from Fig.~\ref{fig:s4} for clarity.

\begin{figure}
    \centering
    \includegraphics{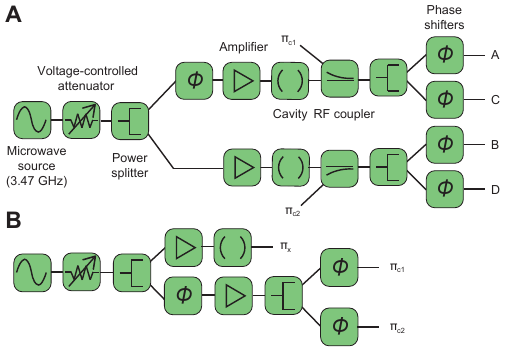}
    \caption{ \textbf{Schematics of the microwave system.} (\textbf{A}), Setup for generating the $\sigma^+$ field and feed for the $\pi$ compensation field. (\textbf{B}), Setup for generating the $\pi$ field. The outputs $\pi_\mathrm{c1}$ and $\pi_\mathrm{c2}$ are feed into the $\sigma+$ path via radio frequency (RF) couplers.}
    \label{fig:s4}
\end{figure}

{\bf Rabi frequencies, microwave ellipticities and orthogonality.}
To measure $\Omega_{\sigma}$, we directly probe the dressed-state structure spectroscopically using methods as described in Ref.~\cite{zhang2024effects}. First, the molecules are adiabatically prepared with single-frequency shielding. Then, a probe field, generated by a loop antenna at the bottom of the vacuum chamber, drives the transition from $\ket{+, \ N}$ to $\ket{-, \ N-1}$, where $N$ is the photon number and $\ket{+}$ ($\ket{-}$) is the upper (lower) dressed state. The Rabi frequency can be found via $\Omega_{\sigma} = \sqrt{(\omega - \omega_0 - \Delta_\sigma)^2 - \Delta_{\sigma}^2}$, where $\omega_0$ is the frequency of the $\ket{0,0}$ to $\ket{1,1}$ transition, $\omega$ is the dressed-state transition frequency, and $\Delta_\sigma$ is the detuning. We note that presence of narrow-band cavity filters prevents direct measurement of Rabi frequencies via $\ket{J = 0}$ to $\ket{J = 1}$ rotational spectroscopy. 

Similarly, $\Omega_\pi$ is measured spectroscopically. Here, the molecules are prepared under double microwave dressing, and again the probe antenna is used to find the transition between the dressed states. With $\Omega_\sigma$, $\Delta_\sigma$, $\Delta_\pi$, $\omega_0$, and the measured transition frequency, $\omega$, known, $\Omega_\pi$ can be obtained from a calculation of the energies in the three-level \textit{V}-system as described in the following section.

The ellipticity of the $\sigma^+$ field us minimized by minimizing the electrostriction of the BEC~\cite{stevenson2025upcoming_1}. We estimate that the residual ellipticity is less than 1${\degree}$.

Compared to our recent BEC work~\cite{bigagli2023observation}, imperfections in the relative alignment between the $\pi$ and $\sigma^+$ have been substantially reduced, which has further enhanced the effectiveness of collisional shielding. Our method to achieve orthogonality between the $\pi$ and $\sigma$ fields is similar to the procedure described in Ref.~\cite{bigagli2023observation}. A residual $\sigma^-$ projection of the $\pi$ field reduces the evaporation efficiency as it directly reduces the shielding quality. Therefore, we  use the molecule number after evaporation as a metric to reduce the $\sigma^-$ projection. By tuning the relative phase and amplitude of $\pi_c$ and optimizing the evaporation quality, we find a combination of parameters that minimizes the $\sigma^-$ projection. Then we use our ability to drive photon changing transitions between dressed states to observe and minimize the $\sigma^+$ projection. As discussed in Ref.~\cite{karman2024upcoming}, when both dressing fields are on, a loss channel opens up, when the resonance condition $n \delta \omega = |E_1 - E_2|$ (or $n \delta \omega = |E_1 - E_3|$) is met, inducing direct transitions to an unshielded state. Here, $E_i$ is the energy of the $i$-th eigenstate (see Eq.~(\ref{eq:mat}) in the next section),  $\delta \omega$ is the frequency difference between the $\pi$ and $\sigma$ fields, and $n$ is an integer. For this calibration, we use $n = 3$. The strength of this resonance is proportional to $|\pi_{\sigma^+}|^2$, where $\pi_{\sigma^+}$ denotes the electric field strength of the unwanted $\sigma^+$ contribution of the $\pi$ field. In practice, we iteratively:
\begin{enumerate}
    \item pick the relative phase and amplitude of $\pi_{c1}$ and $\pi_{c2}$,
    \item minimize $\pi_{\sigma^-}$ with the $\pi_{x}$-$\pi_{c}$ phase and $\pi_{c}$ global amplitude, and
    \item measure $\pi_{\sigma^+}$.
\end{enumerate}
At the end, we pick the $\pi_{c1}$-$\pi_{c2}$ parameters that best minimize $\pi_{\sigma^+}$ and $\pi_{\sigma^-}$. Our estimate is that, in terms of Rabi frequencies, we achieved $|\pi_{\sigma^-}| < \sin{(1^\circ)} |\pi_\pi|$ and $|\pi_{\sigma^+}| < \sin{(4^\circ)} |\pi_\pi|$. The main limitation of this technique is that the optimization is performed for a $\pi$ detuning that is $\approx 2~$MHz larger than for shielding (at that detuning the closest photon-changing transition is found). We use identical cables and cable lengths for the two $\pi$ paths, but there is a small frequency dependent relative phase shift leading to uncertainty in the quality of compensation at the precise dressing frequency.

{\bf Three-level \textit{V}-system and effective induced dipole moment calculation.}
When the molecules are dressed by a $\pi$- and a $\sigma^{+}$-polarized microwave, the single-molecule Hamiltonian under the rotating-wave approximation can be written in the bare basis set of $\{ \ket{0,0},\ket{1,0},\ket{1,1} \}$,
\begin{equation} \label{eq:mat}
\begin{aligned}
& ~~~~ \ket{0,0} ~~~~ \ket{1,0} ~~~~ \ket{1,1}\\
H = \hbar
& \begin{pmatrix}
 0 & \Omega_{\pi}/2 & \Omega_{\sigma}/2\\
\Omega_{\pi}/2 & -\Delta_{\sigma}&0\\
\Omega_{\sigma}/2 & 0  & -\Delta_{\pi}
\end{pmatrix}.
\end{aligned} 
\end{equation}
This Hamiltonian is diagonalized to obtain eigenstates $\{\ket{s},\ket{a},\ket{b}\}$ with eigenenergies  $\{E_s,E_a, E_b\}$. The molecules are prepared in the highest energy dressed dressed state $\ket{s}= \alpha\ket{0,0} + \beta\ket{1,0}+\gamma\ket{1,1}$, which features collisional shielding. This simplified formulation assumes the polarization of the microwave fields to be pure. A detailed theoretical discussion including the ellipicity of both microwave fields is provided in \cite{karman2024upcoming}.

{\bf One-body loss.}
To accurately measure the one-body loss rate $\Gamma_\mathrm{1B}$, we choose data sets with small two- and three-body losses and fit these simultaneously to the loss model with a common one-body lifetime $\tau_\mathrm{1B} = 1/\Gamma_\mathrm{1B}$ . We minimize the total $\chi^2$ found by summing the $\chi^2$ of individual fits. This procedure yields a one-body lifetime of 3.0(5)~s for our data under single microwave dressing (three data sets used). For the data under double microwave dressing, the one-body lifetime is found to be 8.5(1.5)~s (five data sets used). Compared to the single microwave dressing data, for the double microwave dressing the one-body lifetime is improved because phase noise of the microwave system was further reduced using narrow-band filter cavities after the amplifiers. In principle, the one-body lifetime can be extracted by fitting an exponential decay to the number evolution in a sample with very low molecule density. However, the signal-to-noise ratio is low due to small molecule number, leading to a large uncertainty. 

{\bf Coupled-channels calculation.}
We determine two-body loss rates and the positions of bound states in the collisional potential under double microwave dressing from coupled-channels  calculations that are described in detail in Ref.~\cite{karman2024upcoming}. The molecules are modeled as rigid rotors that interact with one another through dipole-dipole interactions and with the two microwave electric fields. We compute adiabatic potential curves by diagonalizing the Hamiltonian excluding radial kinetic energy at fixed $R$, and use these to determine the positions of field-linked bound states by sinc-DVR calculations~\cite{colbert:92} on the adiabatic potential correlating to the initial s-wave channel. To compute two-body loss rate coefficients, we numerically solve the coupled-channels equations subject to an absorbing boundary condition at short range that models collisional loss~\cite{karman2018microwave}. We note that compared to single-field calculations, convergence of these calculations requires larger ``photon number'' basis sets for each of the two fields, ranging from $N_0-4$ to $N_0+2$ compared to $N_0-2$ to $N_0$ in the single-field case, where $N_0$ is some large reference number of photons.
The reason for this is that raising and lowering the photon number in the two fields gives rise to nearly-degenerate channels split only by the frequency difference between the two-fields and inelastic transitions to these channels determines the residual loss rate under double microwave dressing, as is described in detail in Ref.~\cite{karman2024upcoming}.

{\bf Kinetic model.} 
The kinetic model used to fit the number evolution follows the description in Refs.~\cite{bigagli2023collisionally,stevenson2024three}. The model includes one-body, two-body, three-body, and evaporative losses, which can be described by the following equations:
\begin{align}
\dot{N} & = \dot{N}_\text{1B} + \dot{N}_\text{2B} + \dot{N}_\text{3B} + \dot{N}_\text{ev},\\
\dot{E} & =  \dot{E}_\text{1B} + \dot{E}_\text{2B} + \dot{E}_\text{3B} + \dot{E}_\text{ev}+\dot{E}_\text{H}.
\label{eq:de}
\end{align}
Here, $N$ is the molecule number and $E$ is the total energy with $E=3Nk_\mathrm{B}T$. The one-body loss is described by $\dot{N} = - N/\tau_\text{1B}$ and $\dot{E}_\text{1B} = -E/\tau_\text{1B}$. The two-body loss is captured by $\dot{N}_\text{2B} = -\beta_{\mathrm{2B}}n_0 N/(2\sqrt{2}) $ and the related energy change $\dot{E}_\text{2B} = -(3/4) \beta_\text{2B} n_0 E/(2\sqrt{2})$. Here, $n_0 = N \bar{\omega}^3 [ m / (2 \pi k_B T)]^{3/2}$ is the peak density ($\bar{\omega}$ is the mean trap frequency, $m$ is the mass of the molecule). The number loss due to three-body loss is given by  $ \dot{N}_\text{3B}  = -L_\text{3B} n_0^2 N / (3 \sqrt{3}) $ and the corresponding energy change is given by $\dot{E}_\text{3B}  = - (2/3) L_\text{3B} n_0^2 E / (3 \sqrt{3})$.

The effect of evaporative loss is summarized by $\dot{N}_\textrm{ev} = -N \nu(\eta) \Gamma_{\rm th}$~\cite{luiten1996kinetic, olson2013optimizing}. Here, $v(\eta) = (2 + 2\eta + \eta^2)/(2e^\eta)$ is the ratio of elastically scattered molecules with kinetic energy larger than the trap depth,  $\eta = k_\text{B}T/U_\text{min}$ is the truncation parameter, and $\Gamma_\text{th}$ = $n_0 \sigma_{\rm el} v_{\rm th}/N_{\rm col} $ is the thermalization rate. $\sigma_{\rm el}$ is the elastic scattering cross-section, $v_{\rm th} = 4\sqrt{k_{\rm B}T/(\pi M)}$ is the thermal velocity and $N_{\rm col}$ is the number of collisions to produce a $1/e$ change in the molecule temperature. The energy loss due to evaporation is $E = -(1/3) E\alpha(\eta)\Gamma_\text{th}$, where $\alpha(\eta) = (6 + 6\eta + 3\eta ^2 + \eta^3 )/(2e^\eta)$. The full discussion of evaporative loss is given in Ref.~\cite{bigagli2023collisionally}. 

Finally, we include an additional heating term $\dot{E}_\text{H}$. Empirically, we determine that $\dot{E}_\text{H} = 3Nk_{\rm B}\dot{T}_\text{c} + E_\text{n}Nn^2_\text{0}/(3\sqrt{3})$ well describes the experimentally observed increment of the temperature. Here, $\dot{T}_\text{c}$ is a constant one-body heating rate which likely results from mechanical shaking of the dipole trap. This value is fixed to 8(2) nK/s obtained from measuring the temperature evolution of cesium atoms in the same trap. The term $E_\text{n}Nn^2_\text{0}/(3\sqrt{3})$ describes a density dependent heating process parameterized by a fitting parameter $E_\text{n}$. Details on density-dependent heating will be reported in future work~\cite{stevenson2025upcoming}.

{\bf Projected maximum density.} 
The maximum density in Fig.~5 is determined as follows: For a given two-body loss coefficient $\beta_\mathrm{2B}$ and three-body loss coefficient $L_\mathrm{3B}$, the maximum peak density ($n_\mathrm{max}$) for a given overall loss rate $\Gamma = 1/\tau$ can be estimated by solving the following quadratic equation:
\begin{equation}
    \frac{1}{2!}\frac{4}{7}\beta_\mathrm{2B}n_\mathrm{max}+\frac{1}{3!}\frac{8}{21}L_\mathrm{3B}n_\mathrm{max}^2 = \Gamma.
    \label{eq:5}
\end{equation}
We assume a target lifetime of the molecular sample of $\tau = 100$ ms, corresponding to $\Gamma = 10$ 1/s. We also assume that the molecular sample is in the BEC regime, which leads to a suppression factor 1/2! (1/3!) for the two-body (three-body) loss arising from quantum statistics. The factor 4/7 originates from the integral $1/(N n_0) \int{n_\mathrm{TF}}^2d^3r$ and 8/21 from $1/(N n_0^2) \int{n_\mathrm{TF}}^3d^3r$. Here $n_\mathrm{TF}$ is the Thomas-Fermi density profile of a BEC and $n_0$ the peak density. The measurement of three-body losses in this work is performed for temperature of about 100 nK. We project the three-body loss rate coefficient at 10 nK using the semiclassical temperature scaling law $L_{\rm 3B}(T) \sim T^{7/6}$\cite{stevenson2024three}. For parameters in which the measured three-body loss is bounded by the detection limit, we set it to zero. For the two-body loss rate coefficients, in the case of single microwave dressing the experimental values of Fig.~3\textbf{C} are used and in the case of double microwave dressing the theoretical values of Fig.~4\textbf{A}.

% \bibliography{Literature}
%merlin.mbs apsrev4-1.bst 2010-07-25 4.21a (PWD, AO, DPC) hacked
%Control: key (0)
%Control: author (8) initials jnrlst
%Control: editor formatted (1) identically to author
%Control: production of article title (-1) disabled
%Control: page (0) single
%Control: year (1) truncated
%Control: production of eprint (0) enabled
%